# Environmental Adaptability and Mutants: Exploring New Concepts in Particle Transport for Multi-Scale Simulation

M. Augelli, M. Begalli, M. Han, S. Hauf, C. H. Kim, M. Kuster, M. G. Pia, P. Queiroz Filho, L. Quintieri, P. Saracco, H. Seo, D. Souza Santos, G. Weidenspointner and A. Zoglauer

*Abstract*–Ongoing investigations to introduce software techniques suitable to support new experimental requirements for multi-scale simulation are discussed.

## I. INTRODUCTION

Major Monte Carlo transport codes are challenged by new experimental requirements, which have emerged in the recent years. Research in nanodosimetry, nanotechnology-based detectors, radiation effects on components in space and at high luminosity colliders, nuclear power, plasma physics etc. have shown the need of new methodological approaches to radiation transport simulation along with new physics functionality not yet or incompletely covered by general purpose Monte Carlo systems.

A common requirement has emerged in various research domains: the ability of changing the scale at which the problem is treated in the simulation environment. This requirement goes beyond the traditional issues of variance reduction, for which current Monte Carlo codes provide a variety of tools and techniques.

Significant technological developments both in software and computing hardware have also occurred in recent years; they could provide support to the introduction of new concepts in particle transport.

## II. AREAS OF RESEARCH

An R&D (research and development) project [1][2], named NANO5, has been recently launched at the Italian Institute of Nuclear Research (INFN) to address methods in radiation transport simulation and to explore software design techniques suitable to support them. The project was initiated by a small INFN team participating in Geant4 [3][4], that also contributed to the RD44 [5] R&D phase of Geant4; it currently gathers an international team of physicists and engineers with background in high energy physics, astrophysics, nuclear physics and bio-medical disciplines.

The project explores possible solutions to cope with experimental requirements emerged in recent years; it evaluates whether and how they could be supported by Geant4 kernel design.

The main topic of investigation is the issue of performing simulation at different scales in the same experimental environment. This objective is associated with the research of transport methods across the current boundaries of condensed-random-walk and discrete transport schemes. It is also foreseen to explore the possibility of exploiting and extending already existing Geant4 features to apply Monte Carlo and deterministic transport methods in the same simulation environment.

Other issues have been identified along with the experience of Geant4 development and usage over the past years, which would profit from exploratory research in the kernel design:

- Customization of physics modeling in a simulation application
- Scattered and tangled concerns across the code
- Facilities for physics verification and validation
- Computational performance

Recently, a new topic of research has been introduced: the management of epistemic uncertainties [6], i.e. uncertainties related to intrinsic lack of knowledge, in a Monte Carlo system.

The Geant4 toolkit is the ideal playground for this research, thanks to the object oriented technology it adopted in the RD44 phase.

M. Augelli is with CNES, Toulouse, France (e-mail: mauroaugelli@mac.com).

M. G. Pia and P. Saracco are with INFN Sezione di Genova, Via Dodecaneso 33, I-16146 Genova, Italy (telephone: +39 010 3536328, e-mail: MariaGrazia.Pia@ge.infn.it).

L. Quintieri is with INFN Laboratori Nazionali di Frascati, Frascati, Italy (e-mail: Lina.Quintieri@lnf.infn.it).

M. Han, H. Seo and C. H. Kim are with the Department of Nuclear Engineering, Hanyang University, Seoul 133-791, Korea (e-mail: mchan@hanyang.ac.kr; shee@hanyang.ac.kr; chkim@hanyang.ac.kr).

G. Weidenspointner is with MPE, Garching, Germany, and with MPI Halbleiterlabor, Munich, Germany (e-mail: Georg.Weidenspointner@hll.mpg.de).

A. Zoglauer is with the Space Sciences Laboratory, University of California at Berkeley, Berkeley, CA 94720, USA (e-mail:zog@ssl.berkeley.edu).

S. Hauf and M. Kuster are with TU Darmstadt, D-64289 Darmstadt, Germany (e-mail: steffen.hauf@skmail.ikp.physik.tu-darmstadt.de, markus.kuster@xfel.eu).

Marcia Begalli is with State University of Rio de Janeiro, Brazil.

P. Queiroz Filho and D. Souza Santos are with Institute for Radiation Protection and Dosimetry IRD, Rio de Janeiro, Brazil.

## III. ONGOING ACTIVITIES

The activity currently in progress has an exploratory character: it evaluates various problem domains to identify the issues to be addressed [7][8], experimental requirements and candidate technologies.

The ongoing R&D explores concepts and techniques through the development of concrete prototypes. This approach allows one to distill promising candidates for further development among several preliminary ideas and possible technologies. Like in any scientific exploration, not all the investigated candidates will prove worth of further investment; nevertheless, even the negative feedback on some options would be valuable for devising future directions. This attitude characterized the RD44 R&D phase of Geant4.

The project adopts a software process model based on the Unified Process [9] framework, which is use case driven and architecture-centric. The adopted software process involves an iterative and incremental lifecycle.

Regarding multi-scale simulation, the problem domain analysis has identified the concept of "mutability" as a main issue in the context of transition between co-working condensed and discrete transport schemes.

The current research in software design explores the introduction of the concept of "mutants", and of "stimuli" capable of triggering mutations. Related concepts, like reversible and spontaneous mutation, are subject to investigation too.

The introduction of the concept of mutability in physics-related objects requires the identification of their stable and mutable states and behavior, and their fine-grained decomposition into parts capable of evolving, or remaining unchanged.

Two pilot projects are in progress to explore the capability of policy-based class design [10][11] to support this requirement in different physics simulation environments: the environment typical of operation of general-purpose Monte Carlo codes and the environment of so-called "track structure" simulation [12]. In parallel, a project focused on the simulation of radioactive decay [13] explores issues related to the collaboration between electromagnetic and hadronic components of the software design.

Special attention is given to two issues, which could have an impact on experiments of the current generation: the improvement of simulation performance through software design techniques [14] and the facilitation of assessing the reliability of the simulation software.

## IV. CONCLUSION

A R&D project is in progress to investigate the capability of dealing with multi-scale use cases in the same simulation environment. The project is based on Geant4, which offers a versatile playground to explore new concepts and candidate design approaches thanks to its adoption of the object oriented paradigm.

Ongoing exploratory developments are structured as prototypes, which are useful to provide concrete feedback for further consideration of promising solutions, or for discarding options which prove unsustainable.

The results are foreseen to be documented and discussed in depth in dedicated papers.

## ACKNOWLEDGMENT

The authors express their gratitude to CERN for support to the research described in this paper.

The authors thank Sergio Bertolucci, Simone Giani, Vladimir Grichine and Andreas Pfeiffer for valuable discussions.